\documentclass[a4paper,11pt]{article}
\pdfoutput=1 

\usepackage{jinstpub} 
\usepackage{subfig}
\usepackage{graphicx}
\usepackage{hyperref}
\usepackage{chngcntr}
\usepackage{amsmath}

\usepackage{fontenc} 
\usepackage{times} 
\usepackage{mathtools,lipsum}

\title{\boldmath A Sub-Picosecond Digitally-Controlled Phase Delay}

\author{D.~Dehmeshki,}
\author{E.~Frahm,}
\author{R.~Rusack,}
\author{R.~Saradhy,}
\author{Y.~Tousi\note{Corresponding author.}}
\affiliation{The University of Minnesota,\\Minneapolis, Minnesota, USA}

\emailAdd{ymtousi@umn.edu}

\abstract{The use of precision timing measurements will be a major tool at the HL-LHC, where it will be used to suppress pile-up and to search for long-lived particles. To control a reference clock with sub-picosecond accuracy, we have fabricated in the TSMC 65~nm process a digitally controlled phase shifter. It is composed of a chain of 66 cells, each with a digitally controlled planar wave guide with either a short or long delay. With this a reference clock's phase can be controlled to a precision of 200~fs with dynamic range of 12~ps.}


\proceeding{TWEPP 2021\\
  20$^{\text{th}}$ - 24$^{\text{th}}$ September 2021\\
  Virtual}

\begin{document}
\maketitle
\flushbottom

\section{Introduction}
\label{sec:intro}

Precision measurement and control of signal delay is instrumental in both wireless and wireline communication systems. The ability to adjust the delay of a reference clock with sub-picosecond accuracy is necessary to meet the challenge in particle physics of measuring signal arrival times with a sub-picosecond precision. 
To distribute a precisely synchronized reference clock from a single source to different detector elements, which can be separated by distances of up to 20m, requires that subtle variations, such as temperature variations in optical fibers, can be detected and corrected in real time. A core part of a calibration scheme is a high precision time delay that can track and adjust for such variations. 
True time delays have been in use since the early days of microwave systems for tuning a variety of components and signals. The traditional method for delay control is based on physical adjustment of a transmission line using a trombone-like mechanical knob \cite{opticaldelay}. Such a method can provide a reliable and accurate control over the delay of phase across a broad frequency. While trombone type delay lines are widely used in smaller and discrete component systems, they have limited use in more complex integrated systems. This is mainly due to the relatively large size of the structure limiting the number of tunable delays in the system, as well as its slow and manual means of delay adjustment. 

A large body of literature is dedicated to improving on these limitations by replacing mechanical tuning with electrical switches or analog control. A common method to replace mechanical adjustment of the length is to do so by switching the signal path between two predesignated paths, each with its own unit length \cite{switched_delay,trombone_delay}. These approaches rely on artificial construction of a true-time delay that is subsequently adjusted by switching in and out different sections. A cascaded series of such switched delays with either binary or uniform delay steps can ideally provide accurate tuning of the effective length similar to the trombone structure. However, due to the need for multiple switches in series to the signal path such systems require low loss electrical switches to avoid significant signal degradation. Diode-based switches have demonstrated the feasibly of such methods in off-chip implementations. However, due to the higher loss in standard integrated technologies, such as CMOS, such signal switching is not effective for on chip implementation.


The objective of this work is to develop a true time delay line structure that is fully integrated, while maintaining the benefit of traditional tunable delays. The remainder of this paper will discuss the proposed structure, followed by simulation results and prototype implementation and measurement results.  

\section{Tunable transmission line}
The relation between the input and output of a TEM mode transmission line closely represents an ideal delay line. This is because it represents wave propagation in a passive broadband structure that provides a dispersion-free and broadband group delay. Although physical adjustment of the length has proven to a be powerful delay tuning technique, an on-chip implementation of a transmission line imposes a fixed overall length. Furthermore, multi-path switched-based length adjustment has limited utility, due to their limited accuracy and insertion loss. However, due to its linear dispersion relation, the concept of using a transmission line is superior to multi-section phase adjustment, and it is highly desirable to implement a tunable delay structure than maintains the core benefit of a dispersion free transmission line. 

The output of a loss-less and dispersion-free transmission line, shown in Fig. \ref{fig1}, can be related to its input by the following equation:
\vspace{-0.05in}
\begin{equation}
v_o\left(t\right)=v_i\left(t-t_d\right),
\end{equation}
where, $t_d=L$/$V$ is the effective delay of this transmission line, $L$ is the length of the line and $V$ is propagation velocity of the wave in the line. This suggests that in addition to adjusting the length, the delay can also be controlled by adjusting the propagation velocity, which depends on the speed of light in the material, given by $V=c/\sqrt{\epsilon\mu}$. As it is very difficult to adjust these material properties in an normal transmission line, an artificial transmission line can be constructed as shown in Fig. \ref{fig1} using lumped components, which can mimic the behavior of a transmission line with a highly granular structure.
\subsection*{Artificial transmission line}
Given the linear nature of this delay line, one can divide it into small unit delays represented by lumped circuit elements, such as inductors and capacitors. This is theoretically possible for any linear electromagnetic system as long as individual unit cells have a small phase delay compared to the wavelength \cite{pozar2011microwave}. In such a representation, we can divide the transmission line into $N$ unit delay sections, each consisting of an inductance $L$ and capacitance $C$. Solving for the relation between the input and output voltage in this equivalent circuit, the relationship between the continuous wave parameters and the circuit values in this artificial construct can be found. In each section the delay line the characteristic impedance is $Z_0=\sqrt{L/C}$ and the delay is $\sqrt{LC}$. By combining $N$ cells a transmission line with a total delay of $t_d=N\sqrt{LC}$ can be constructed. 
 \begin{figure}[htbp]
 \centering 
 \includegraphics[width=.8\textwidth]{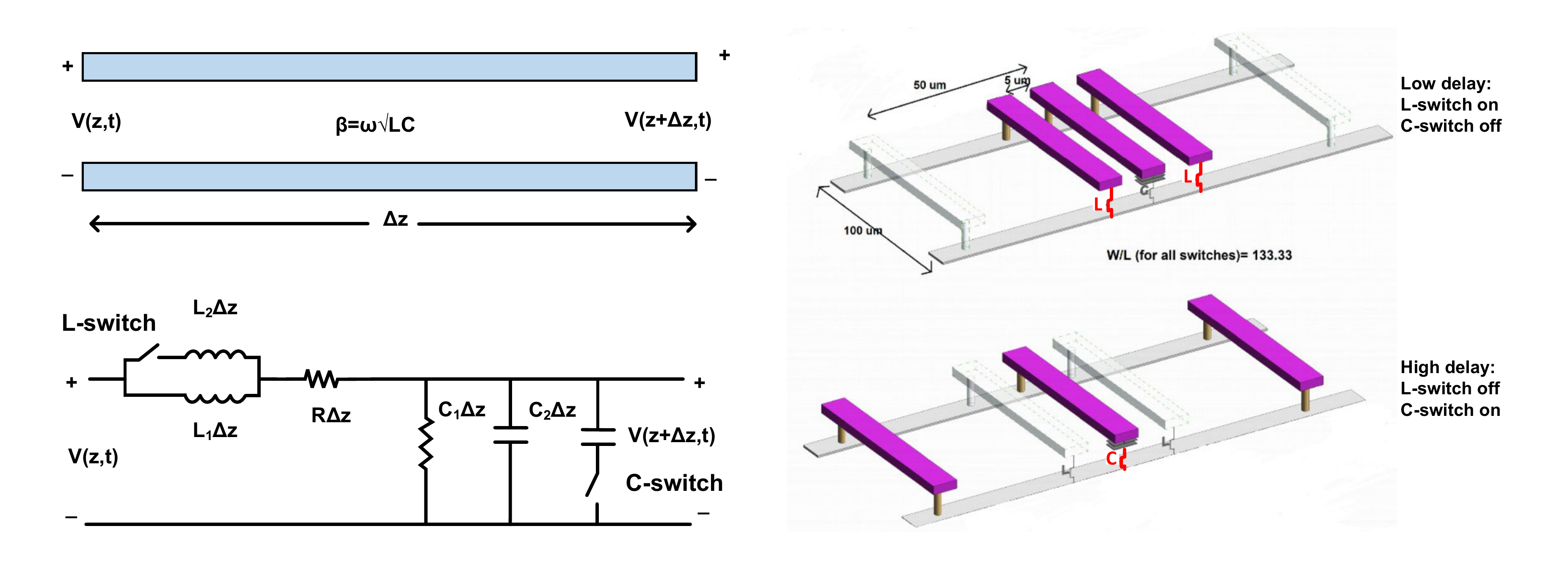}
\   \caption{Left: An tunable transmission line and its equivalent distributed circuit with tunable values. Right: physical layout of the delay unit cell and the two delay settings.}\label{fig1}
\end{figure}

This representation is only valid when delay per section is small compared to the signal period, that is:
\begin{equation}
    f\ll\frac{1}{\sqrt{LC}}.
\end{equation}
This criteria needs to be taken into account in the design of the delay and the choice of the physical size of the unit cells, which will depend on the maximum frequency of operation required for the delay line. 

\begin{figure}[htbp]
 \centering 
 \includegraphics[width=.9\textwidth]{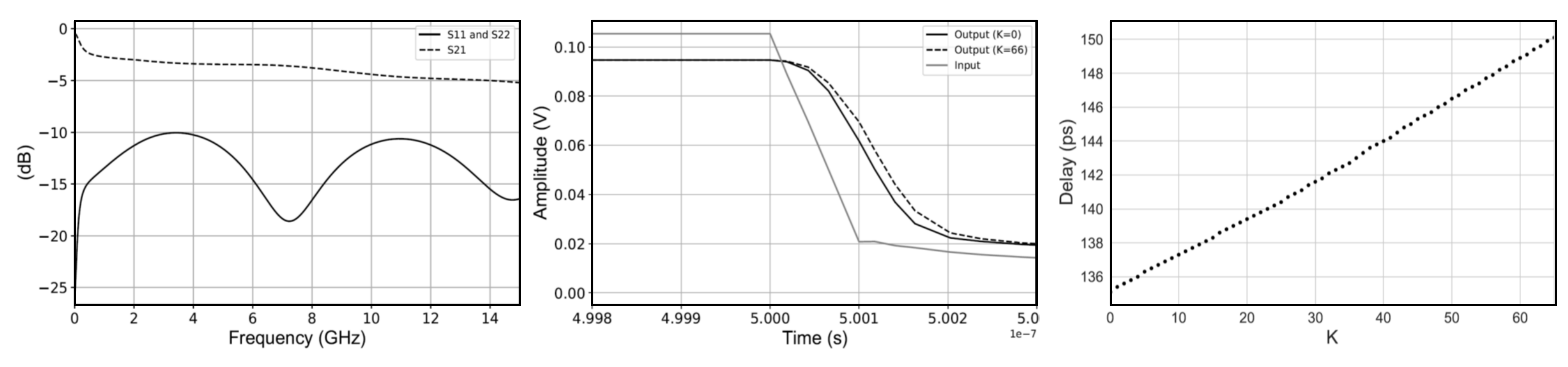}
\   \caption{Simulation results of the proposed delay line: left: S-parameters, middle: pulse waveforms, right: pulsed input delay.}\label{fig2}
\end{figure}

The overall delay is adjusted by controlling the the unit delay of each cell. To do this we introduce switched inductors and capacitors to the delay cell. As shown in Fig. \ref{fig1}, each cell consists of two switches. The L-switch adjusts the ground return path of the transmission line: when the switch is on the return path is close to the signal line resulting in a smaller magnetic flux and inductance compared to the off mode. Similarly, the C-switch adds a shunt capacitor to the unit structure. The combination of these two switches allows control of both the delay, $\sqrt{LC}$, and $Z_o$. This allows the delay to be adjusted without modifying the impedance of the line and affecting wave propagation. This effectively results in a dispersion free transmission line with a tunable propagation velocity. Furthermore, due to the granular nature of this control, this method can provide a highly accurate control of the delay by simply breaking down the transmission line into small unit cells.
Simulation results of the group delay for different settings is shown in in Fig. \ref{fig2}.

The dimensions of the parallel lines and the value of the characteristic impedance are chosen based on electromagnetic extraction of the distributed structure, followed by a Spice based simulation of the extracted structure, along with the lumped capacitances and switches. In addition to minimal dispersion, a delay step of significantly less than 1~ps can be achieved. Due to the inherently discrete tuning method in this digitally-controlled delay line, the size of the unit delay cell plays a critical role in the overall resolution of the structure. As a result, we designed the unit cells with the smallest possible dimensions, while allowing sufficient margin within the fabrication design rules. Fig. \ref{fig1} shows the dimensions of one of the 66 delay cells used in this design.

In order to adjust the delay line, a serial interface provides access to the  L-bit and C-bit in each delay cell. To adjust the delay, the design employs a thermometer coding scheme where the delay cells are switched in a sequential manner, starting from all the cells being in the low delay mode and subsequently switching them one-by-one to the high-delay mode. This results in a gradual increase in the total delay, while minimizing abrupt transitions in the characteristic impedance of the line. As a result, the delay steps are monotonic and uniform.

\subsection*{Chip implementation}
The chip prototype is fabricated using the TSMC 65nm CMOS process. The chip consists of two parallel delay lines for use in a differential signal environment. As shown in the die micrograph, the 66 cells are placed in three rows of 22 cells for an area efficient design. The 180-degree bend in the structure required between adjacent rows is optimized to create equal delay with minimal effect on the line dispersion. We use the top metal for both ground and signal routing, while the switches are placed as close as physically possible to the return path to ensure efficiency. We measured the chip using the Keysight N5242A PNA-X Network Analyzer with the two sides of the delay line connecting to the PNA through differential RF probes.
 \begin{figure}[htbp]
 \centering 
 \includegraphics[width=.9\textwidth]{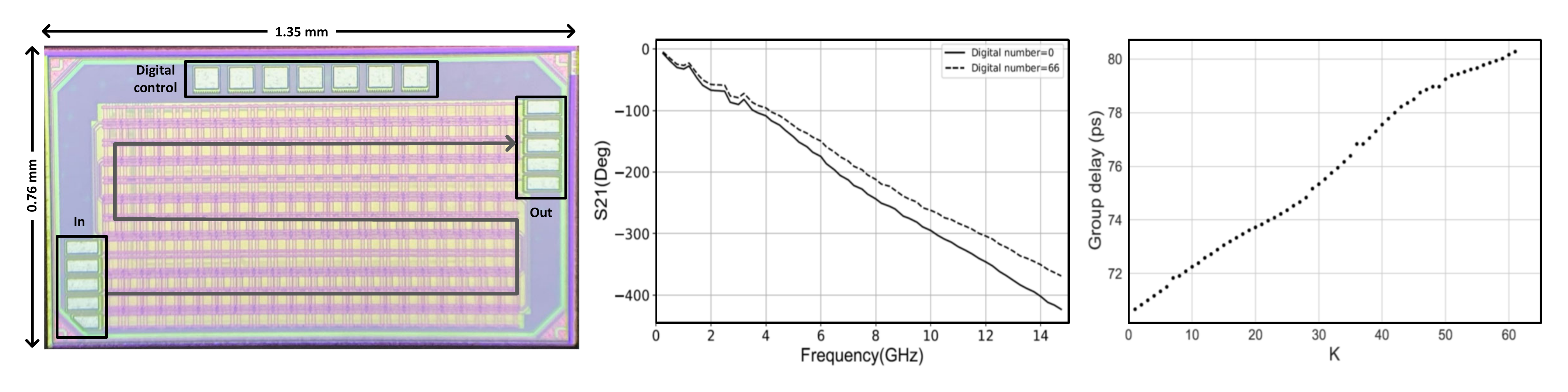}
\   \caption{Chip level measurements. Left: chip microphoto, middle: phase vs. frequency for minimum and maximum settings, group delay vs. digital setting. }\label{fig3}
\end{figure}

Fig. \ref{fig3} illustrates the measured phase S-parameters of the line from DC to 15GHz, showing the delay vs. frequency characteristic for the highest and lowest delay settings of the circuit. The measured average group delay and loss for all 66 digital codes demonstrates the linear delay vs. digital relationship. The tunable delay line provides a total delay range of 9.4psec with an average delay step of 140fs. The maximum phase error is less than 0.15 degrees and there is a close alignment between simulation and measured S-parameter data.

\section{Board-level testing}

\begin{figure}[ht]
\begin{minipage}[t]{0.35\linewidth}
    \centering
    \includegraphics[width=0.7\linewidth]{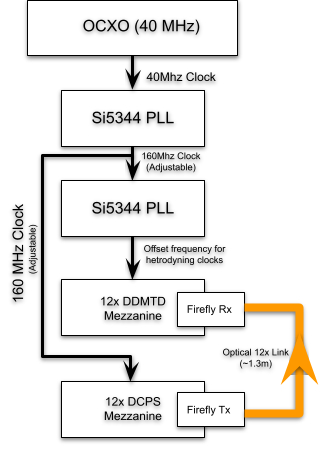}
    \end{minipage}
\begin{minipage}[t]{0.55\linewidth}
    \centering
    \includegraphics[width=0.9\linewidth]{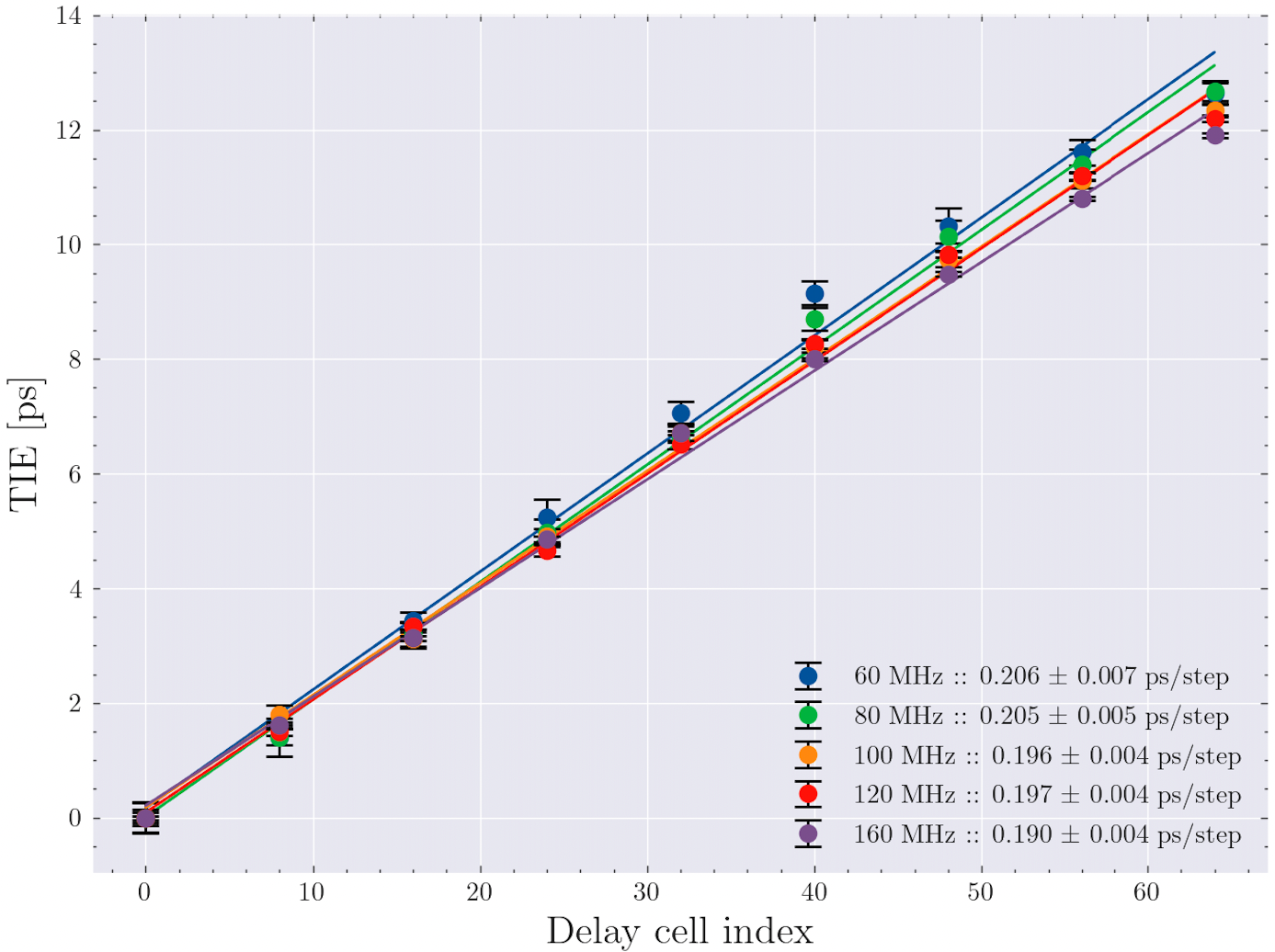}
\end{minipage}

\caption{Left: The test setup used to measure the changes of phase in the time domain. Right: The phase changes measured with the DDMTD for digital clock frequencies between 60 and 160 MHZ.}
\label{fig:phase-shifts}
\end{figure}

As a further test of the structure and as a demonstration of its use with digital signals, we have tested the delay line with a digital clock at different clock frequencies. For this we have used a circuit similar to the one described in our previous presentation at TWEPP 2019 \cite{TWEPP2019}. Our test setup is sketched in Fig. \ref{fig:phase-shifts}, in it the output of an oven-controlled crystal oscillator with a frequency of 40 MHz is input to Silicon Labs PLL (Si5344) to generate a highly stable clock at selected frequencies. This clock is then injected into the circuit and the phase is controlled using the serial interface.
The output of the ASIC is coupled to a Samtec Firefly 12TX transmitter, which is coupled through a multimode fiber optic cable to a Samtec Firefly 12RX receiver on the same PCB. The received signal is compared with the original clock signal using a digital dual mean time difference circuit. 

The results obtained at different clock frequencies for different settings of the phase delay are shown in Fig. \ref{fig:phase-shifts}, and the step size is measured to be $\approx200$fs.

\vspace {-0.2in}
\section{Summary}
\vspace{-0.05in}
The proposed transmission-line based solution enables substantially better delay resolution across a wide bandwidth, while ensuring a monotonic response. Furthermore, due to the passive nature of the line, the proposed delay line is more energy efficient compared to active delay tuning methods. As a result, this chip demonstrates a robust tunable true-time delay with sub-picosecond accuracy for precision timing applications.

\vspace{-0.05in}
\acknowledgments
\vspace{-0.05in}
We are grateful the US DOE Office of High Energy Physics for their support under the award DE-SC0020185.


\bibliographystyle{JHEP.bst}
\bibliography{ref.bib}




\end{document}